\documentstyle[psfig]{mn}
\newif\ifAMStwofonts

\newcommand{\mic}{$\mu$m}

\title[Template near-IR galaxy spectra]
  {Near-infrared template spectra of normal galaxies: 
  k-corrections, galaxy models and stellar populations.}

\author[Mannucci et al.]
{
F. Mannucci$^1$, 
F. Basile$^2$, 
B.M. Poggianti$^3$, 
A. Cimatti$^4$, 
E. Daddi$^2$,
\newauthor
L. Pozzetti$^5$, and 
L. Vanzi$^6$\\
$^1$ C.A.I.S.M.I. - C.N.R., Largo E. Fermi 5, I-50125, Firenze \\
$^2$ Universit\`a di Firenze, Largo E. Fermi 5, I-50125, Firenze \\
$^3$ Osservatorio Astronomico di Padova, vicolo dell'osservatorio 5,
I-35122 Padova \\
$^4$ Osservatorio Astrofisico di Arcetri, Largo E. Fermi 5, I-50125, Firenze \\
$^5$ Osservatorio Astronomico di Bologna, via Ranzani 1, I-40127, Bologna\\
$^6$ European Southern Observatory, Alonso de Cordoba 3107, Santiago,
Chile\\
}

\date {Submitted. Accepted}
\pubyear{2001}

\begin{document}

\maketitle


\begin{abstract}

We have observed 28 local galaxies in the wavelength
range between 1 and 2.4 $\mu$m in order to define template spectra of 
the normal galaxies along the Hubble sequence. 
Five galaxies per morphological type were observed in most cases, and
the resulting RMS spread of the normalized spectra of each class, 
including both intrinsic
differences and observational uncertainties,
is about 1\% in K, 2\% in H and 3\% in J. Many absorption
features can be accurately measured.
The target galaxies and the spectroscopic aperture
(7\arcsec$\times$53\arcsec) were chosen to be
similar to those used by Kinney et al. (1996) to define template
UV and optical spectra. The two data sets are matched in order to
build representative spectra between 0.1 and 2.4 $\mu$m.
The continuum shape of the optical spectra and the relative
normalization of the near-IR ones were set to fit the average
effective colours of the galaxies of the various Hubble classes. 
The resulting spectra are used to compute the k-corrections of the
normal galaxies in the near-IR bands and to check the predictions of
various spectral synthesis models: while the shape of the continuum is
generally well predicted, large discrepancies are found in the
absorption lines. 
Among the other possible applications, here we also
show how these spectra can be used to place constraints on the dominant 
stellar population in local galaxies.
Spectra and k-corrections are publicly available and can be downloaded
from the web site http://www.arcetri.astro.it/$\sim$filippo/spectra.
\end{abstract}


\section{Introduction}

Galaxies are complex systems where gas, dust, stars and dark matter
deeply interact with each other for a very long time. The emerging
spectrum is therefore sensitive to a number of parameters as mass, age,
metallicity, star formation history, dust, geometry etc. 
Despite this complexity, many groups have tried to model the galaxy
emission, and the observed optical spectra can now be
accurately reproduced 
(e.g. Bruzual \& Charlot, 1993; Worthey, 1994; Fioc \& Rocca-Volmerange, 1997;
Barbaro \& Poggianti, 1997; Tantalo et al., 1996).
This is increasingly important as more high-redshift galaxies are discovered 
and important quantities as age, mass and star formation history
are estimated by comparing the integrated properties of the galaxies
with the predictions of spectrophotometric models.

These models are calibrated on local galaxies and the first test on
their properties is the comparison of the modeled spectra
with the observed ones over a large wavelength range.
Until now, this comparison could
only be done in the optical and near-UV region where representative 
galaxy spectra are available
(e.g., Pence, 1976; Coleman, Wu \& Weedman, 1980; Kennicutt, 1992a, 1992b; 
Kinney et al., 1996, hereafter K96), 
 while photometric points are used for all the other wavelengths.
The aim of this work is to extend the range where 
a full comparison is possible towards longer wavelengths, between 1 and 2.4
$\mu$m. This is in particular important because large
differences are present between the different models especially in this
wavelength range where the relative contribution of dwarf, giant and 
supergiant stars is poorly known (e.g., Charlot, Worthey \& Bressan
1996).
These stellar populations are characterized both by different depths of the
absorption lines and different shapes of the near-IR continuum
(see, for example, Pickles, 1998; Lan\c con et al., 1999) and therefore 
by studying breaks, colours and lines of a full spectrum it is possible, in
principle, to separate their contribution.

Near-IR spectra can also be used
to compute the k-corrections of an unevolving galaxy population, and, 
therefore, to study the luminosity functions and the 
differential evolution of the galaxies
(Pence, 1976; Coleman et al. 1980, Frei \& Gunn 1994, Poggianti, 1997).
The near-IR k-corrections are becoming more and more important as new
IR-optimized telescopes are becoming available, as the Very Large
Telescope, the Large Binocular Telescope and the Next Generation Space
Telescope.
Complete spectra between UV and near-IR wavelength can also play a role
in the measure of the redshifts of large samples of galaxies,
both from photometric data (for a review see Yee 1998) 
and even from spectroscopic data (eg., Glazebrook, Offer
and Deeley, 1998).

In this paper we show the results of the observations of 28 galaxies
of morphological type between E and Sc, while later classes 
will be the subject of a future work. Target selection, observations and
data reduction are described in sec.~\ref{sec:obs}. Great care was taken 
in matching the near-IR spectra to the optical ones by using the
observed UV-optical-IR colours, as described in sec.~\ref{sec:calib} and
sec.~\ref{sec:colors}.
In sec.~\ref{sec:kcorr} we derive the k-corrections in the J, H and K
band. In the following sections, we compare the total
spectra to the predictions of some spectrophotometric models for elliptical
galaxies and use the observed absorption lines to study the dominant
stellar populations.
Spectra and k-corrections are available in electronic form.

\section{Observations and data reduction}
\label{sec:obs}

The target galaxies were selected in the morphological classes
E, S0, Sa, Sb and Sc. In each class we selected large nearby galaxies without
indication for peculiar activity. An absolute magnitude threshold of
M$_V<-21$ was used to ensure a homogeneous metallicity and
to allow a sample as close as possible to that in K96.
We also mainly observed galaxies whose optical spectra were available in the 
literature to be also able to study single objects.
Therefore we selected galaxies 
from the samples by K96 and Kennicutt (1992a, 1992b),
adding one more elliptical
galaxy from the list given in Goudfrooij (1994). The resulting 28 targets
are listed in 
Table \ref{tab:objlist}, together with the morphological informations
derived either from the LEDA catalog (http://leda.univ-lyon1.fr/leda/)
or from the reference listed in the table. The objects are sorted in the
table and divided into 5 classes according to the morphological index T
(see, for example, Buta et al., 1994).

{\tabcolsep 1.6mm
\begin{table}
\begin{center}
\caption{Observed galaxies 
\label{tab:objlist}}
\begin{tabular}{lllllcc} 
\hline
\hline
NGC &\multicolumn{2}{c}{R.A.~~~(1950)~~~DEC.}&Type&T$^a$&Ref.$^b$&Bands$^c$\\ 
\hline
\multicolumn{6}{c}{E ($-5\le$T$<$-3) }\\
\hline
4648 & 12:39:54.5 & +74:41:44 & E3   & -4.9 & 1 & JHK \\ 
1700 & 04:54:28.1 &--04:56:30 & E4   & -4.8 & 2 & JHK \\ 
3379 & 10:45:11.3 & +12:50:48 & E2   & -4.8 & 1 & JHK \\ 
4472 & 12:27:13.9 & +08:16:22 & E1   & -4.7 & 1 & JHK \\ 
4889 & 12:57:43.7 & +28:14:54 & E4   & -4.3 & 1 & JHK \\ 
\hline
\multicolumn{6}{c}{S0 ($-3\le$T$<$0) }\\
\hline
1023 & 02:37:15.5 & +38:50:56 & E-S0 & -2.6 & 3 & JHK \\ 
3245 & 10:24:30.1 & +28:45:45 & S0   & -2.1 & 1 & JHK \\
4350 & 12:21:25.1 & +16:58:21 & S0   & -1.8 & 3 & JHK \\
4382 & 12:22:52.8 & +18:27:59 & S0-a & -1.3 & 3 & JHK \\ 
5866 & 15:05:07.0 & +55:57:20 & S0-a & -1.3 & 1 & JHK \\ 
\hline
\multicolumn{6}{c}{Sa (0$\le$T$<$2) }\\
\hline
2681 & 08:49:58.0 & +51:30:14 & S0-a &  0.4 & 3 & JHK \\ 
3623 & 11:16:18.6 & +13:22:00 & SBa  &  1.0 & 1 & JHK \\ 
2775 & 09:07:41.0 & +07:14:35 & Sab  &  1.7 & 1 & JHK \\ 
3368 & 10:44:06.9 & +12:05:05 & SBab &  1.7 & 1 & JHK \\ 
\hline
\multicolumn{6}{c}{Sb (2$\le$T$<$4) }\\
\hline
4826 & 12:54:16.9 & +21:57:18 & Sab  &  2.4 & 3 & JHK \\ 
4736 & 12:48:31.9 & +41:23:32 & Sb   &  2.5 & 3 & JHK \\ 
2841 & 09:18:34.9 & +51:11:19 & Sb   &  3.0 & 3 & ~~HK \\
4102 & 12:03:51.6 & +52:59:23 & SBb  &  3.1 & 3 & JHK \\ 
3147 & 10:12:39.3 & +73:39:02 & Sbc  &  3.7 & 1 & JH~~ \\ 
\hline
\multicolumn{6}{c}{Sc (4$\le$T$<$6) }\\
\hline
2903 & 09:29:20.2 & +21:43:19 & SBbc &  4.0 & 1 & JHK \\ 
5194 & 13:27:45.9 & +47:27:12 & Sbc  &  4.1 & 3 & JHK \\
3994 & 11:55:02.3 & +32:33:23 & Sc   &  4.9 & 3 & JHK \\ 
1637 & 04:38:57.5 &--02:57:11 & Sc   &  5.0 & 3 & ~~HK \\
2276 & 07:10:22.0 & +85:50:58 & Sc   &  5.4 & 1 & ~~~~K \\
\hline
\multicolumn{6}{c}{Others}\\
\hline
3432 & 10:49:42.9 & +36:53:09 & Irr  &  9.0 & 3 &  \\ 
2798 & 09:14:09.4 & +42:12:34 & Sap  &  1.7 & 1 &  \\
1569 & 04:26:05.8 & +64:44:18 & Irr  &  8.2 & 1 &  \\ 
4449 & 12:25:45.9 & +44:22:16 & Irr  &  9.3 & 1 &  \\ 
\hline
\end{tabular}
\end{center}
$^a$: morphology stage index (e.g., Buta et al., 1994)\\
$^b$: references. 
1:~Kennicutt (1992b);  2:~Goudfroij (1994); 3:~K96. \\
$^c$: bands used for the templates
\end{table}
}

We planned to have 5 galaxies for each class between E and Sc with 
good S/N ratio (SNR) to have an estimate of the intrinsic spread in the spectra 
of each class and to average out peculiar characteristics. 
This number was not reached for the Sa class because NGC2798
turned out to be peculiar (it is part of an interacting system)
and with emission lines.  This galaxy is listed in the ``Others''
section of Table~\ref{tab:objlist} together with 3 more observed objects 
of later stages not used for the templates.

\begin{figure}
\centerline{\psfig{figure=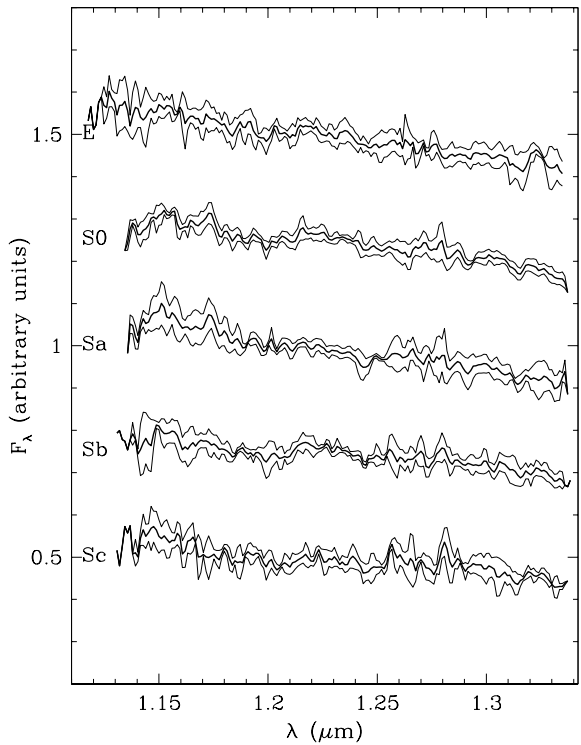,width=8cm}}
\caption{Rest frame average spectra of each class of galaxies in the J band. 
The thick line is the average of the observed spectra, the thin lines
show the ranges within 1 standard deviation. 
Arbitrary offsets were added to the spectra for clarity.}
\label{fig:spectraJ}
\end{figure}

\begin{figure}
\centerline{\psfig{figure=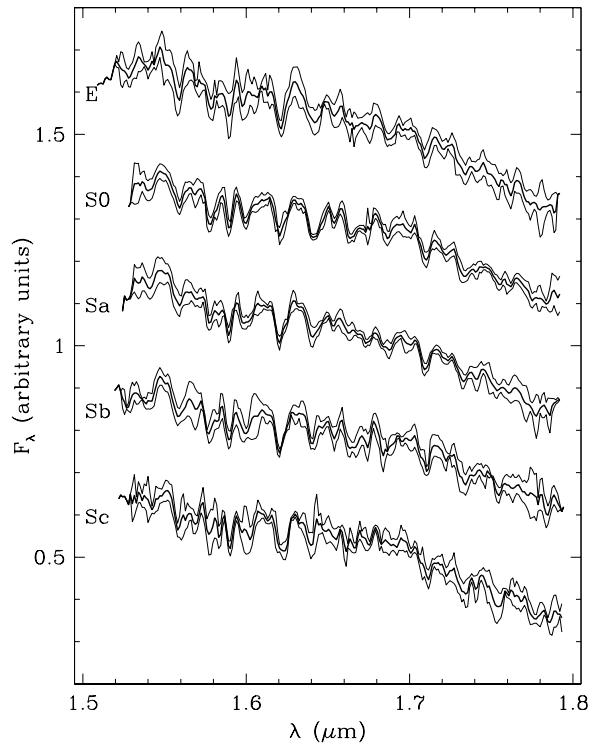,width=8cm}}
\caption{As Figure~1, H band}
\label{fig:spectraH}
\end{figure}

\begin{figure}
\centerline{\psfig{figure=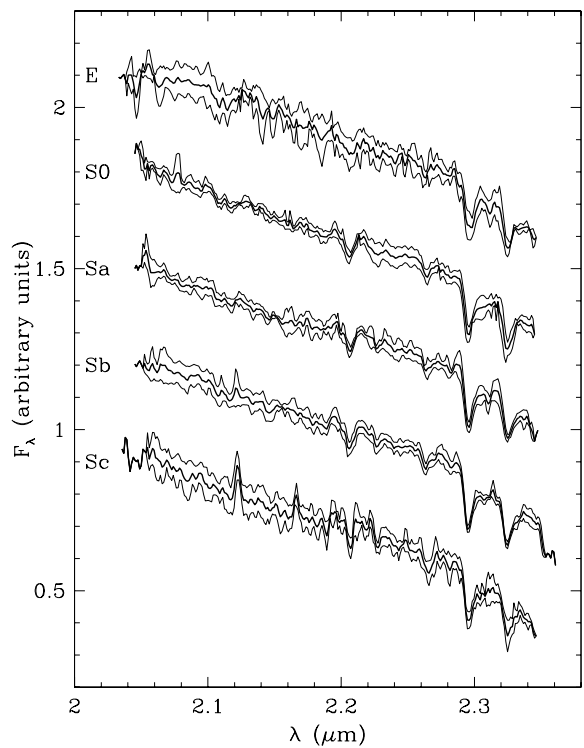,width=8cm}}
\caption{As Figure~1, K band}
\label{fig:spectraK}
\end{figure}

All the spectra were obtained at TIRGO, an IR-optimized
1.5m telescope on the Swiss Alps. We used the long-slit 
spectrometer LonGSp (Vanzi et al., 1997b) which is based on a NICMOS3
256$\times$256 pixels Rockwell array and is sensitive between 0.9 and
2.5 \mic. 

Each galaxy was observed through the largest available slit, 7\arcsec\
wide, in order to have a good coverage of the galaxy.
Spectra were obtained in the
J, H and K bands for integration times between 15 and 30 minutes for
each band. 
The resulting spectral resolution is about 300 in J, 400 in H and 500 in K.
The telescope was chopped every minute between the
galaxy and the nearby sky to have a good sky subtraction. Each time the galaxy
was placed on a different position of the slit to minimize the effect of bad
pixels and distortions of the flat-field. For each galaxy, stars of 
spectral type between F8 and G8 were observed in order to measure 
the atmospheric transmission and the variation of the instrumental efficiency 
with the wavelength, as discussed in Maiolino \& Rieke (1996). 
The total number of acquired images is about 7000. 

All the spectra were sky subtracted using the average of the adjacent 
spectra, and divided by differential flat-fields obtained by 
imaging the dome. The resulting frames
were then rectified, co-aligned and stacked together using a clipping algorithm.
A one-dimensional spectrum was then extracted from
the central 53$\arcsec$\ in order to maximize the observed fraction of the 
galaxies. 
The resulting aperture of 7\arcsec$\times$53\arcsec\ 
is similar to the aperture of 10\arcsec$\times$20\arcsec\ used by K96 
to define their templates and allows us to merge the two data sets.
At the average distance of the galaxies of our sample, 27$h^{-1}_{50}$
Mpc ($h_{50}=$H$_0/50$ km/sec/Mpc), 
the spectroscopic aperture corresponds to about 1$\times$7 Kpc.
We estimated the fraction of the galaxy light collected by the slit
by using the existing aperture photometry in the V band 
(Prugniel and H\`eraudeau, 1998): on average, our spectra contain about
20\% of the galaxy light, ranging from 7\% for the closest and largest
galaxies to about 35\% for the smallest ones. 
Therefore our results should not be considered ``total'' spectra, even if
the difference is expected to be negligible in the near-IR (see discussion
below). This is more important when also the K96 spectra are considered
as in the optical the radial gradients are expected to be larger (but see also 
the discussion in K96).

\begin{figure}
\centerline{\psfig{figure=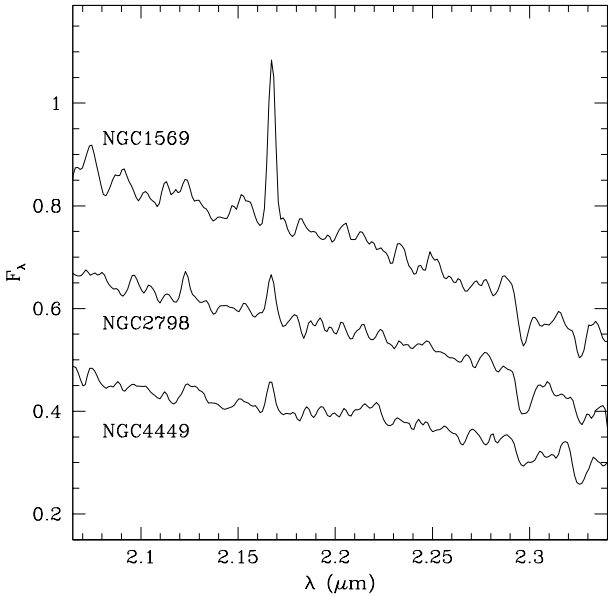,width=9cm}}
\caption{K-band spectra of the galaxies with active star formation 
NGC1569, NGC2798 and NGC4449, The emission lines at 2.122$\mu$m (H2) 
and 2.166 (Br$\gamma$) can be seen in the spectra.}
\label{fig:active}
\end{figure}

The ``true'' spectra of the reference stars
were modeled by the solar spectrum (Livingston \& Wallace, 1991) 
corrected for the slightly different temperature of the stars. 
The derived transmission curve was then applied to the galaxy spectra. 
The final spectrum of each galaxy in each band is then 
corrected for Galactic extinction, normalized, shifted to zero
redshift, and averaged with
the other galaxies of the same class. A clipping algorithm was applied
to remove a few residual discrepant value due to  cosmic rays,
erratic bad pixels not present in the mask and uncorrected features of the
reference stars. 
In Figures~\ref{fig:spectraJ},\ref{fig:spectraH} and \ref{fig:spectraK} 
we show the average spectra and the relative
standard deviations in the three bands. 

The differences between the spectra of the galaxies of the same class
is due to both all the observational
uncertainties (noise, residual sky contribution, 
distortion on the flat field, non perfect correction of the instrumental and
atmospheric transmission) and the intrinsic differences between the observed
galaxies. We can therefore have a robust upper limit 
of the total uncertainties
of our spectra by looking at the resulting RMS among the galaxies of the
same type. This quantity is between 0.7\% and 1.4\% in K, 
between 1.7\% and 2.4\% in H 
and between 2.4\% and 3.6\% in J, depending on the morphological class. 
The tendency to have larger spreads towards bluer wavelengths 
is probably dominated by larger intrinsic differences, and the value in K, 
about 1\%, can be assumed as the real uncertainty in the single spectrum.
Because of the different
redshift of the galaxies, less galaxies contribute to the edges of the
spectra and therefore the uncertainties tend to be larger in the
external part (about 5\%) of the bands. The parts of the
spectra without indication of standard deviation in figures 1,2 and 3
have been derived from only a galaxy and should be treated with caution.

If the normal galaxies show a high degree of uniformity,
on the contrary large differences are found for the few galaxies
with peculiar properties or of later types. The K spectra of 3 of these
galaxies (NGC1569, NGC2798 and NGC4449) are shown in
Figure~\ref{fig:active}, while for NGC3432 the SNR is too
low. The galaxies in the figure show prominent emission lines whose
brightness and ratios change much from one object to the other. In this
case no attempt was made to create a template.

\section{Calibration of the spectra}
\label{sec:calib}

We now want to establish the relative calibration of the spectra in the 
various bands
to be able to study the overall spectral energy distribution
from the UV to the near-IR. 
It is not possible to derive these relative normalizations 
by simply using spectral observations because of several effects, 
such as slit losses, variability of the 
atmospheric transmission and uncertainties on the
reference star colours. 
The usual method to normalize spectra taken in different bands
is to use the broad-band colours of the target objects: 
spectra should be calibrated to
reproduce the colours of the observed galaxies inside 
the spectroscopic aperture. 

As this work is aimed at the construction of spectra representative of the
average properties of the different classes of galaxies, 
we have calibrated the spectra by using the average colours 
of each class. This method guarantees that the final spectra closely match
the average properties of the class and allows the reduction of the 
uncertainties on the colours as hundreds of galaxies can be used.

{\tabcolsep 1.5mm
\begin{table}
\begin{center}
\caption{Average effective colours of galaxies with M$_V<$-21 
For each colour the standard deviation and the number of used objects
are also reported.
\label{tab:colors}}
\begin{tabular}{cccccccc}
\hline
\hline
       &  U-B  &  B-V  &  V-R  &  V-I  &  V-K  & J-H$^a$& H-K$^a$\\
\hline
    E  &  0.50 &  0.99 &  0.59 &  1.22 &  3.30 &  0.66 &  0.21  \\
       & (0.08)& (0.05)& (0.05)& (0.07)& (0.09)& (0.05)& (0.02) \\
       &  323  &  418  &  314  &  221  &    32 &   225 &  225   \\
       &       &       &       &       &       &       &        \\
    S0 &  0.47 &  0.97 &  0.58 &  1.20 &  3.25 &  0.66 &  0.22  \\
       & (0.11)& (0.08)& (0.05)& (0.08)& (0.14)& (0.05)& (0.02) \\
       &  287  &  344  &  227  &  158  &    13 &   235 &  235   \\
       &       &       &       &       &       &       &        \\
    Sa &  0.36 &  0.90 &  0.58 &  1.17 &  3.24 &  0.67 &  0.25  \\
       & (0.19)& (0.11)& (0.08)& (0.11)& (0.18)& (0.06)& (0.03) \\
       &  138  &  185  &   73  &    82 &    17 &   105 &  105   \\
       &       &       &       &       &       &       &        \\
    Sb &  0.22 &  0.82 &  0.57 &  1.16 &  3.21 &  0.66 &  0.25  \\
       & (0.20)& (0.12)& (0.09)& (0.11)& (0.28)& (0.06)& (0.03) \\
       &   321 &  541  &  156  &  315  &    16 &    93 &   93   \\
       &       &       &       &       &       &       &        \\
    Sc &  0.06 &  0.70 &  0.52 &  1.15 &  3.03 &  0.66 &  0.25  \\
       & (0.18)& (0.13)& (0.10)& (0.15)& (0.24)& (0.07)& (0.04) \\
       &   294 &  536  &  133  &  287  &    23 &    46 &   46   \\
       &       &       &       &       &       &       &        \\
Sd$^b$ & -0.12 &  0.62 &  0.47 &  1.09 &  2.95 &  0.65 &  0.23  \\
       & (0.16)& (0.18)& (0.13)& (0.19)& (0.32)& (0.08)& (0.05) \\
       &   53  &   99  &  25   &  58   &   12  &   26  &   24   \\
       &       &       &       &       &       &       &        \\
 I$^c$ & -0.15 &  0.51 &  0.40 &  1.08 &  2.35 &  0.51 &  0.21  \\
       & (0.20)& (0.17)& (0.20)& (0.30)& (0.35)& (0.10)& (0.06) \\
       & 102   &  117  &  28   &   35  &    5  &   22  &   20   \\
\hline
\end{tabular}
\end{center}
$^a$: The J-H and H-K colours are based also on the results in Fioc
\& Rocca-Volmerange, 1999, where only average quantities are given. 
In these cases the scatter is not measured but estimated.\\
$^b$: M$_V<$-20\\
$^c$: no magnitude selection.
\end{table}
}

Of course, colours inside our rectangular spectroscopic aperture
are not available. However,
the dependence of near-IR colours and spectra with aperture is known to be
very weak because the dominant stellar population at these wavelength
is always the giant branch (with the exception of starburst galaxies, see for
example, Maraston (1998))
and only the second-order metallicity and age gradients could be seen.
Therefore we can use colours inside circular
apertures as they are the only ones usually published, and the difference
in shape between photometric and spectroscopic apertures is
negligible if the two areas are chosen to be similar.
The most similar circular aperture with enough published
photometric data is the ``effective'' one,
i.e., the aperture containing half of the light of the galaxy in the B band.
Effective colours are available for a large number of galaxies and in 
many filters and is therefore possible to have large homogeneous data
sets.

For the smallest objects, the effective aperture is similar to 
the spectroscopic
one, for the largest galaxies the photometric aperture is larger. 
We have performed two checks to be sure that, even in the latter case,
effective colours can reliably be used to calibrate the spectra:
first we have compared spectra extracted from the central 10\arcsec\ of
the slit with those of the outer part; 
second, we have compared the spectra of large and
small galaxies of the same class (as NGC4350 having D$_0$=2\arcmin.0 
and NGC4382 with D$_0$=8\arcmin.2) . 
In both cases no systematic differences are seen.

Once the aperture is chosen, we have also to select the luminosity of
the galaxies used to derive average colours.
As described in the previous section, the target objects are large,
metal rich galaxies, and we want to use galaxies with the same
properties to calibrate their spectra. In fact, smaller galaxies
tend to be less metallic and therefore bluer in the optical colours (see,
e.g., Fioc \& Rocca-Volmerange, 1999), giving origin to the well-known 
colour-magnitude relation. We will use optical and near-IR  colours 
appropriate for galaxies with M$_V<$-21, corresponding to our
sample. 

\begin{figure}
\centerline{\psfig{figure=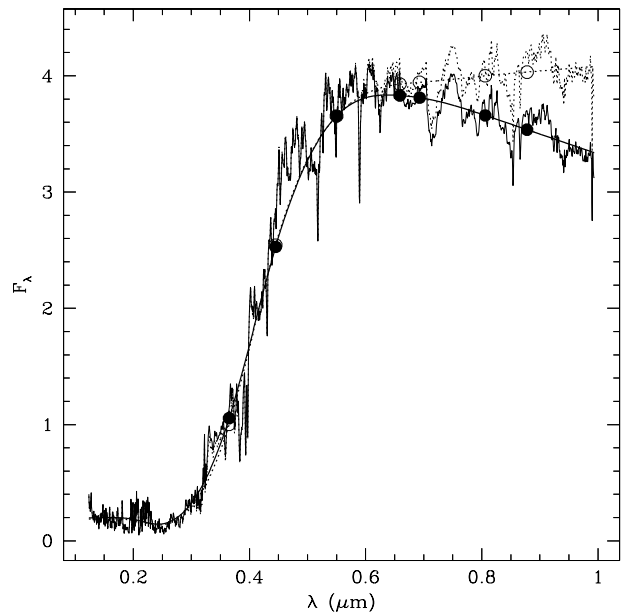,width=9cm}}
\caption{Colour correction to the K96 spectra.
The dotted lines show the K96 spectrum for the
ellipticals and a spline fit to the corresponding photometric points.
(empty dots). 
The solid dots
show the effective colours of the galaxies of this class with M$_V<-21$
(see Table~\ref{tab:colors}), the solid lines are the fit to these data
and the final spectrum, corrected to fit the colours.
\label{fig:kinney}
}
\end{figure}

Many photometry papers where used to derive these colours and check the results
(Aaronson, 1978; Frogel et al., 1978; Persson et al., 1979; 
Griersmith et al., 1982; Glass, 1984; Giovanardi \& Hunt, 1988;
de Jong \& van der Kruit, 1984; de Vaucouleurs et al., 1991; 
Buta et al., 1994, 1995a, 1995b; Prugniel \&
H\`eraudeau, 1998, Fioc \& Rocca-Volmerange, 1999).
The optical colours are mainly based on the  Prugniel \&
H\`eraudeau (1998) catalog after correcting for Galactic dust 
using the RC3 value of the galactic extinction in B 
(de Vaucouleurs et al., 1991) and the
Cardelli et al. (1989) extinction curve. The E and S0 optical-to-IR
colours are based on Persson et al. (1979) and Frogel el al. (1978),
while the colours of later spirals are mainly derived from 
Griersmith et al. (1982), de Jong \& van der Kruit (1984) and Aaronson
(1978).
IR-to-IR colours are based on Glass (1984), Frogel et al. (1978) and 
Fioc \& Rocca-Volmerange (1999). Note that the J-H and H-K colours of
the early-spiral galaxies are redder than those of the ellipticals: a
short discussion about this unexpected effect can be found in 
Fioc \& Rocca-Volmerange (1999).
All the colours were converted to 
Johnson's U, B and V, Cousins' R and I and CIT J, H and K.
Great care was used in comparing different galaxy samples: whenever
possible, the results from different samples were compared to each other
to check for consistency and to discover any important selection effect. 
The results are listed in Table \ref{tab:colors} together with
the colour scatter and the number of galaxies used. The absolute magnitude
threshold was reduced to M$_V$=-20 for the Sd class (6$<$T$<$8) because these
galaxies are on average intrinsically fainter than the other classes, 
and for the
I class (T$>$8) no magnitude restriction was used because of the
large scatter observed. The colours of these latter two classes (not
used for the templates and reported for completeness) 
are based on a smaller number of galaxies and
are probably less accurate that the others.

Earlier classes tend to be more homogeneous than the later ones. For the
ellipticals, the spread of the colour distribution, when removing a few very
discrepant objects, is comparable to the uncertainties on the single
measure as quoted in the original papers, leaving little
room to an intrinsic spread of the population due, for example, to different
metallicity or star formation histories. On the contrary, for the later class
the observed spread is several times larger than the measured uncertainties.
As an example, for the ellipticals the V-K colour distribution has a standard
deviation of 0.09 mag (and an error on the mean of the order of 1\%), 
similar to the average estimated error on the 
single point of about 0.10 mag, while for Sc the observed spread is 0.24
mag.

\section{Matching optical and near-IR spectra}
\label{sec:colors}

\begin{figure*}
\centerline{\psfig{figure=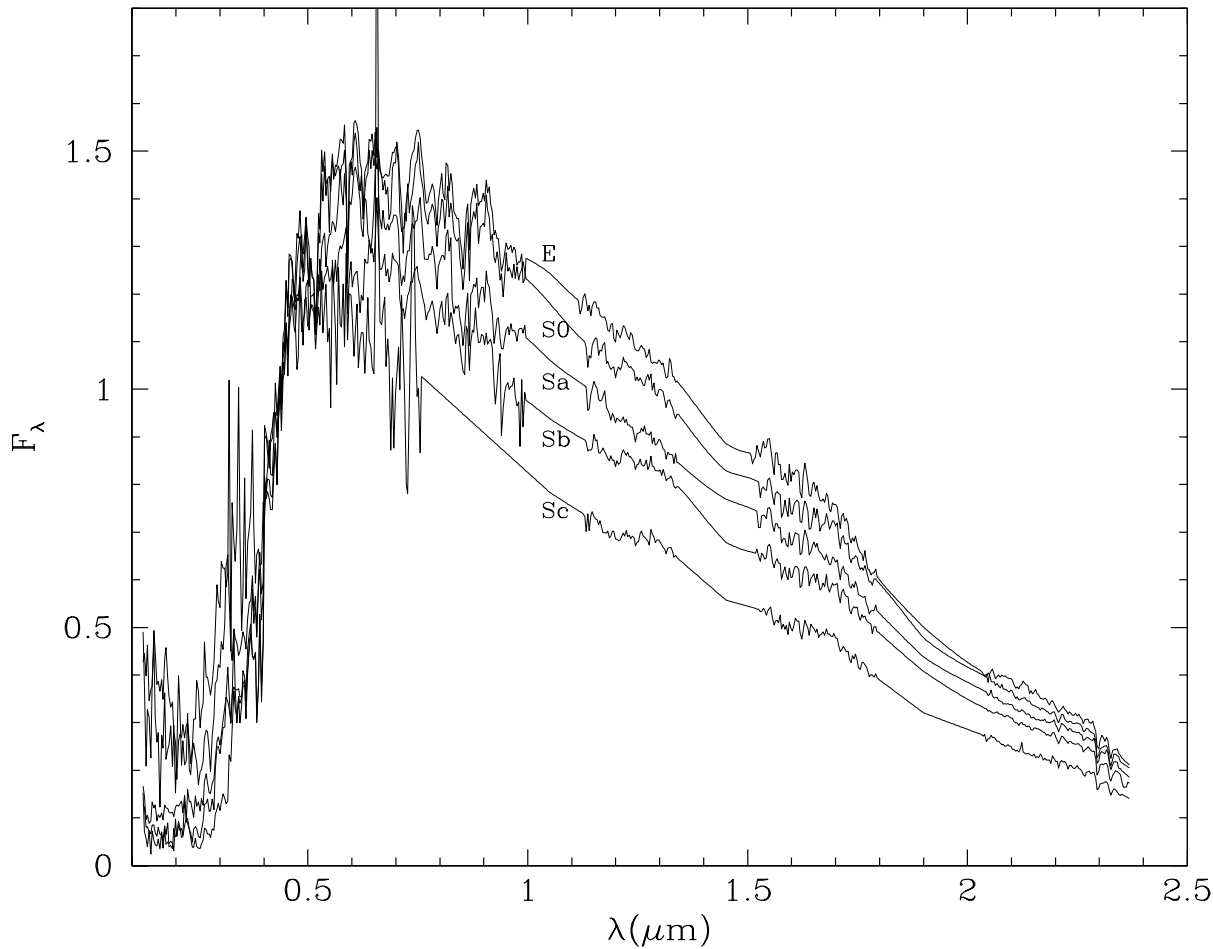,width=17cm,angle=-90}}
\caption{Template spectra for E, S0, Sa, Sb and Sc galaxies. The
spectra were reduced to a resolution of 40\AA\ and normalized to the B band. 
The interpolation between the bands is arbitrary.} 
\label{fig:plotall}
\end{figure*}

The same colour calibration was used to match our spectra to existing
optical and UV ones to obtain a single template from 0.1 to 2.4~\mic.
Several optical templates are available in the literature
(Coleman et al., 1980; Kennicutt, 1992a, 1992b; K96).
In order to obtain meaningful results, the optical spectra
must have appropriate resolution and wavelength coverage,
should be based on a relatively large number of galaxies 
and the used spectroscopic aperture must be similar to our one.
Among the available optical templates, only the K96 spectra
have the required features:
1- they have a high enough resolution (between 6 and 10\AA)
for most studies of stellar populations;  
2- have the largest wavelength coverage, extending from 1200~\AA\ up to 
about 1\mic, at the edge of the J band (the Sc spectrum covers
only up to 7700\AA); 
3- use an aperture of 10$\times$20\arcsec\ at all the wavelengths, 
similar to ours.
4- are based on the observations of 30 galaxies, 12 of which in common
to our sample.

While below 7000\AA\ the K96 spectra closely reproduce the observed colours 
of Table~\ref{tab:colors}, above this wavelength the K96 spectra are 
much redder.
The difference is not due to the chosen colours (effective aperture and
M$_V<-21$) because the K96 spectra 
don't fit the average colours inside any aperture and for any magnitude
(see, for example, Goudfrooij, 1994). Moreover,
they don't reproduce the 
colours of the observed galaxies: as an example, all
4 elliptical galaxies in K96
have observed Johnson's V-I colours between 1.38 and 1.45 for apertures
between 10 and 20\arcsec\ (Prugniel \& H\`eraudeau, 1998), 
while the colour derived from the K96 spectrum is 1.59. 
It should also be noted that above 7000\AA\ the K96 spectra are quite 
different from those by Coleman et al. (1980) who also corrected 
the spectra for the observed colours. No useful comparison can be done 
with the Kennicutt (1992a, 1992b) atlas because it only covers the
spectra region below 7000\AA.

In order to have a self-consistent 
template we have to apply some corrections to the optical spectra to
make them fit the colours.
The procedure is shown in Fig.
\ref{fig:kinney} where the K96 spectrum for the ellipticals is plotted against
the colours of Table \ref{tab:colors} used both to calibrate the
near-IR spectra and to correct the K96 templates. 
The correction to the K96 spectra was computed by 
the ratio of two splines, fitted respectively to the observed colours and to 
the colours derived from the spectra (see Figure~\ref{fig:kinney}). 
This correction is not unique, but the availability of data 
in many filters maintains this uncertainty 
below the other ones.

For all the K96 spectra between E and Sb, a good agreement
(within a few percent) is present below 7000\AA\ between the optical 
spectra and the average colours, and the corrections are small.
On the contrary,
the Sc spectrum, based on two galaxies only, appears to be much bluer than the 
expected colours of its class even at $\lambda<7000$~\AA. 
Also the equivalent widths of the emission lines
are much larger than in the galaxies of our sample, 
about 296\AA\ for the H$\alpha$+[NII] vs. an average of 41\AA\ for our
galaxies (Kennicutt \& Kent, 1983) and 54\AA\ for the
Sc sample by Kennicutt (1992a). 
We think that both these effects are
probably due to the fact that the two Sc galaxies in the K96 sample, 
NGC~598 and NGC~2403, are of
later type (T=5.6 and 5.7) then the average Sc (T=5.0, from Buta et al., 1994)
and those in our sample (T=4.7). 
To correct for this, we have both applied the colour correction, as for
the other types, and also scaled down the emission line to have
the same EW of our sample. 
The resulting spectrum, although representative of an
average Sc galaxy, should be
used with caution when comparing optical to near-IR features and when
dealing with the emission lines, also because of its lower SNR.

As final step, the spectra were arbitrarily interpolated across the wavelength
ranges where the atmosphere is not transparent.
In Figure~\ref{fig:plotall} we show at a reduced resolution the final spectra
reproducing the observed effective colours. 

The relatively small apertures used for both spectroscopy and photometry 
means that the spectra in Figure~\ref{fig:plotall} are representative 
of the central regions of the galaxies containing about half of the
total emission. K96 compare their optical spectra with those by
Kennicutt (1992b) obtained using a much larger aperture (about
90$\times$90\arcsec) 
and conclude that for E, S0 and Sa the K96 spectra are in fact 
representative of the total emission of the galaxies because their integrated 
properties are dominated by the central bulge. The same argument
applies in the near-IR too, and in this case a weaker dependence of the
spectra from the aperture is also predicted: in fact, while in the
optical the young population in the upper main sequence can easily
dominate, in the near-IR any shortly living super-giants population
must compete with the usually much more numerous giants population and 
therefore dominates only in
case of strong activity of star formation.
As discussed in the sec.~\ref{sec:calib}, this can be checked
comparing near-IR spectra extracted from the inner and outer regions of 
the galaxies, or from large and small objects. 
A weaker variation of the near-IR spectra with aperture with
respect to the optical spectra can also be predicted from the colours:
the J-H and H-K colours are only 
weakly dependent of the galaxy type (see Table~\ref{tab:colors}) and
do not show significant radial gradients (Glass 1984; Fioc \&
Rocca-Volmerange, 1999), not even for
the late type galaxies. Therefore we conclude that for the early type galaxies
the spectra in Figure~\ref{fig:plotall} can also be used as templates of the
integrated galaxy. For Sb and Sc galaxy, younger extra nuclear populations 
can change the spectra significantly, both in the optical because of the stars
of the upper main sequence and in the near-IR if an important population of red
supergiants is present.

\begin{figure}
\centerline{\psfig{figure=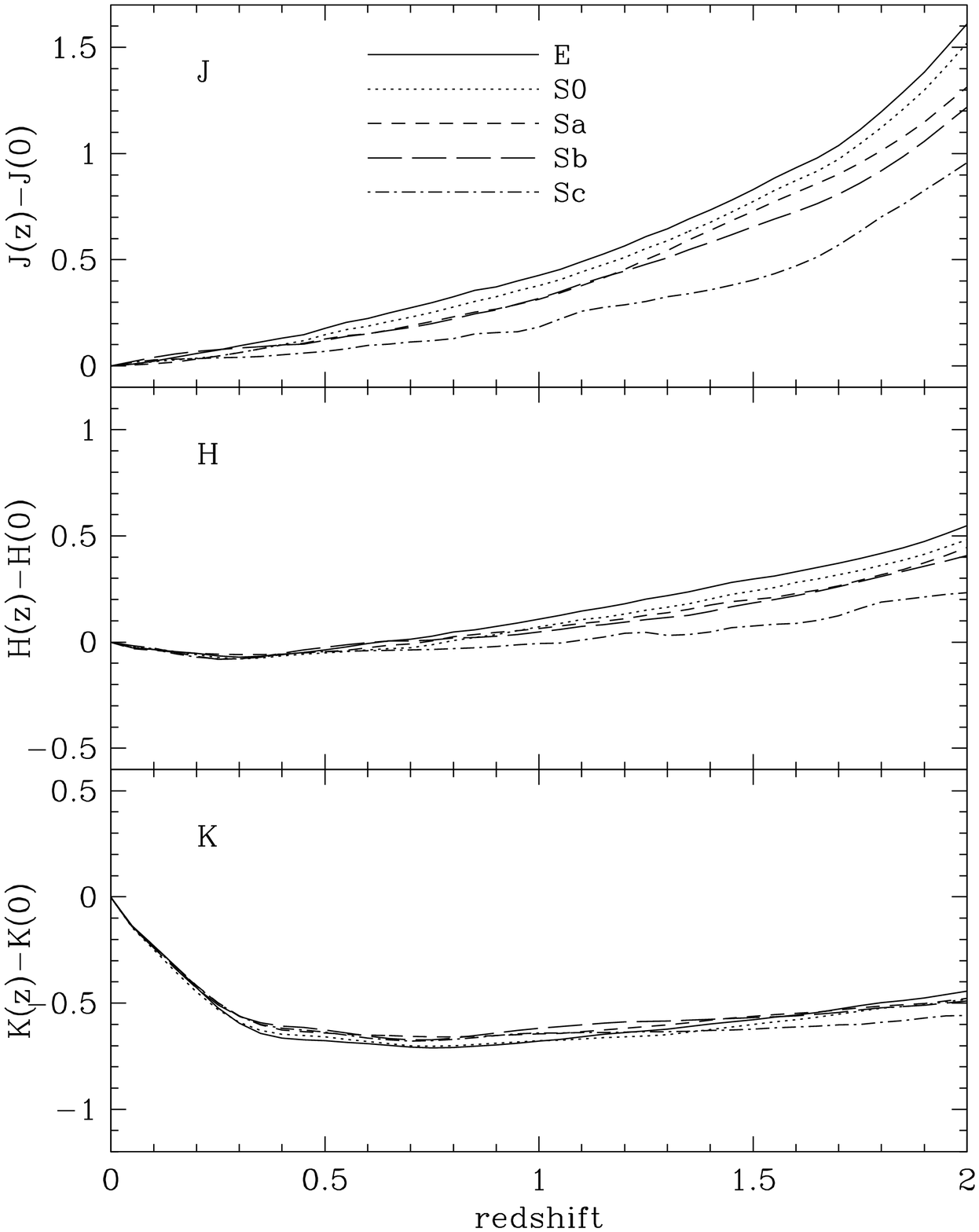,width=9cm}}
\caption{k-corrections in the J, H and K bands  between
redshifts 0 and 2 for each morphological type: E (solid line), S0 (dotted), Sa
(short dashed), Sb (long dashed) and Sc (dot-dashed). Our spectra
define these k-corrections up to z=0.3 in J, z=0.7 in H and 1.2 in K,
while at higher redshift they reflect the K96 spectra corrected to match
the observed colours.}
\label{fig:kcorr}
\end{figure}

\section{The k-corrections}
\label{sec:kcorr}

The first quantities that can be derived from these spectra are 
k-corrections, i.e., how magnitudes and colours 
change with the redshift 
because of the shift of the  observed wavelength ranges.
These corrections are crucial to link the properties of the local universe 
with the distant one (see, for example, Peebles, 1993), and their
accurate knowledge is important to study the evolutionary effects. 
Our spectra allow us to compute these corrections in the near-IR filters 
starting from z=0 with a greater accuracy than by using 
simple broad-band photometry as, for example, Cowie et al. (1994).
In Figure~\ref{fig:kcorr} we show the k-corrections computed for the J, H 
and K bands.
These filters have an
increasing importance in the surveys for distant galaxies and in the
study of the known high-redshift objects, especially when
more 8-meters class space- and
ground-based telescopes optimized for the infrared will be available.

The k-corrections were computed using Bessell \& Brett (1988) filters,
more standard that the CIT system.
We note that in the K band the k-corrections
are almost independent from the galaxy type up to z=2, 
reflecting the dominance of giant stars in these wavelengths in all
the observed galaxies, as previously noted (Glazebrook et al. 1995). 
Note that no contributions are included from galaxy evolution or 
increasing absorption from the intergalactic medium; the computation is
limited to z=2 because above this limit these effects are certainly
dominant.

We have compared our k-correction in the K band with others previously
published.  At low redshift ($z<0.6$) we found consistency  
with the k-correction by Glazebrook et al. (1995) 
derived from Bruzual \& Charlot (1993) model for a SSP with age of 5 Gyrs,
and used, for example, by Loveday (2000) and Szokoly et al. (1998)
to define the local luminosity function.
On the contrary, big differences (up to 0.2 mag) 
are found with k-correction from `UV-hot' elliptical 
model of Rocca-Volmerange \& Guiderdoni (1988).
At higher redshift ($z>0.6$) our k-corrections are `redder' up to 0.3
magnitudes than previous ones, 
however at these redshifts they should be taken with caution,
and considered as lower limits,
due to the possible evolutionary effect on galaxy spectrum.

Important differences are also found with the k-corrections by 
Poggianti (1997). Regarding the SED of the elliptical,
the difference in J is generally about 0.1
and in H is always below 0.2, while differences greater than 0.3 
are found in the K band in the redshift range z=0.3 to 1.3,
up to a value of about 0.45 magnitudes at z=0.4-0.7.
Although the differences in the filter response functions
adopted here and in Poggianti (1997) contribute to the differences between
the K-corrections, most of the K-band discrepancy 
arises from intrinsic differences in the SED, probably due 
to the sparse sampling of the Poggianti (1997) models at $\lambda > 17000$ \AA.
The comparison for Sa and Sc galaxies is less meaningful
because of large spectral variations within each morphological
class, possible aperture effects (see \S4), especially
for Sc galaxies, and the presence of emission lines, not included
in the Poggianti (1997) model.

\section{Comparing observations and models}

Template spectra can be used to test the
spectrophotometric models of the galaxies. Despite their low resolution,
the SNR is high enough to allow useful comparison of both the continuum
shape and the spectral features.
Here we show two examples of the possible use of these spectra. 
The first one, described in this section,
is the comparison of the shape of the continuum
over a large wavelength range. In the next section the absorption lines
will be compared.

Two reasons suggest to start the continuum comparison from the 
elliptical galaxies: first, from the
observational point of view this class is the most homogeneous,
its average colours are better defined and the spectrum has
smaller uncertainties. Second, elliptical galaxies are known to be
dominated by old stellar populations and therefore the
uncertainties on the star formation history (SFH) have lower
consequences than in the case of the spirals: while in the latter case 
detailed SFH are required to fit the spectra, for the ellipticals a
single parameter, the age, is usually enough to obtain a good agreement.

\begin{figure}
\centerline{\psfig{figure=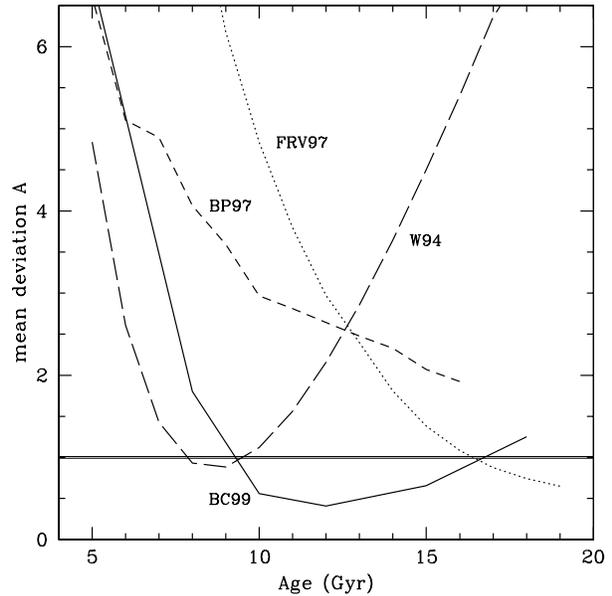,width=9cm}}
\caption{Mean deviation of the predictions of some spectrophotometric
models from the observed spectrum of the elliptical galaxies as a
function of the age of the stellar population. \
Details about the models can be found in the text.
}
\label{fig:fit}
\end{figure}

\begin{figure*}
\centerline{\psfig{figure=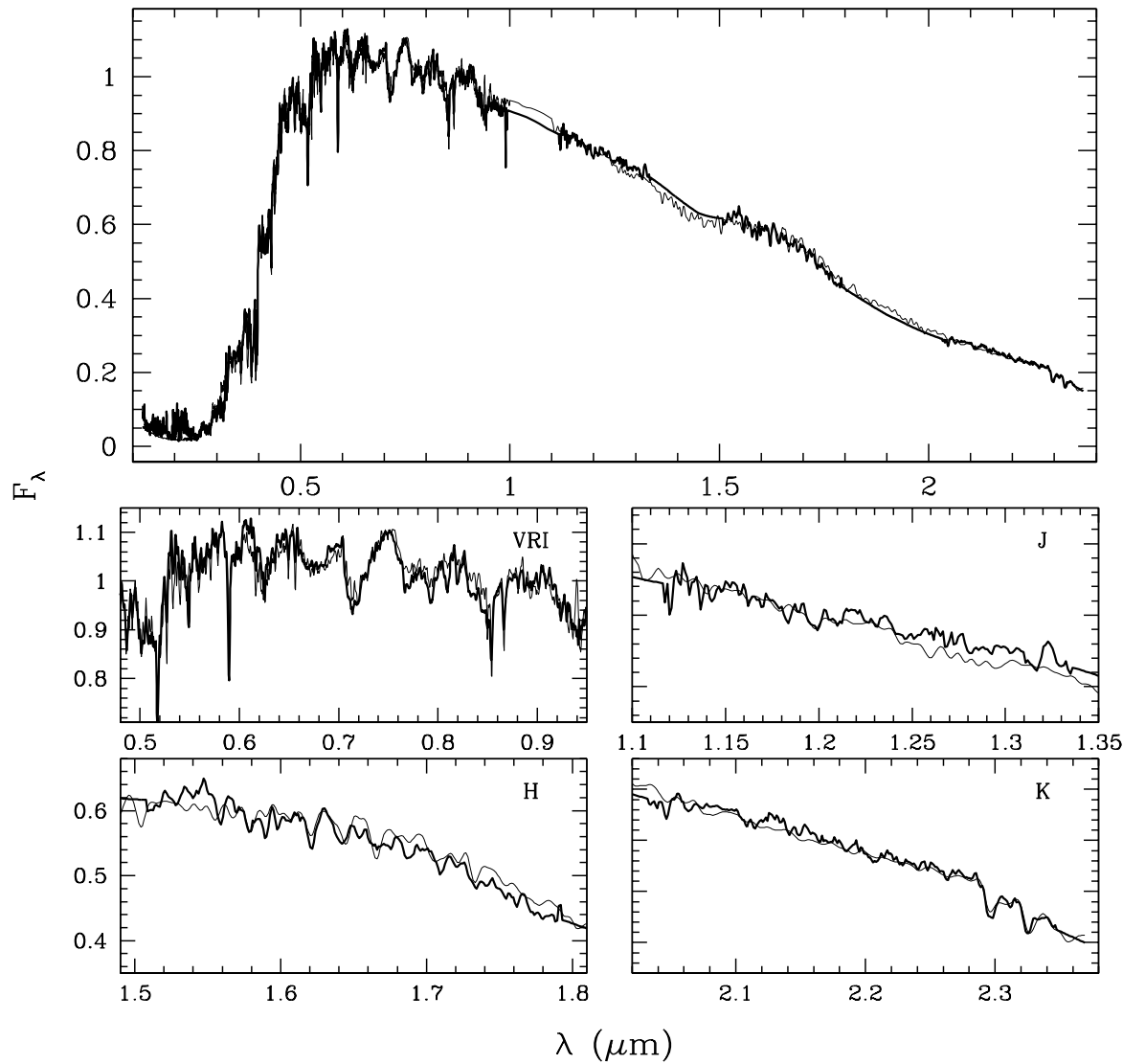,width=16cm}}
\caption{The observed spectrum of the elliptical galaxies (thick line) 
is compared with a SSP spectrum of the BC99 model
(thin line) with an age of 12 Gyr and solar metallicity. 
The upper panel shows the full spectrum, the four lower panels show
enlargements in the optical and in the three near-IR bands.
The shape of the spectrum is very well reproduced
in all the range between 0.2 and 2.4 $\mu$m, and this is true for the
optical absorption features too. In the near-IR large discrepancies can
be noted, especially in J and H. The reason of these differences are
explained in the text.}
\label{fig:bc99}
\end{figure*}

We used four different models: the 
Bruzual \& Charlot model in the 1999 version (BC99), 
the Fioc \& Rocca-Volmerange (1997) PEGASE model (FRV97) 
the Barbaro \& Poggianti (1997) one (BP97), and the model by Worthey
(1994) (W94). 
We have also used a previous (1996) version of the Bruzual \& Charlot model 
(BC96) that allows for non-solar metallicities.

BC99 and BC96 use isochrones from the Padova group (Bressan et al., 1993)
and two different stellar libraries, the empirical library from Pickles
(1999) and
the theoretical one by Lejeune et al. (1997). The latter library permits
to predict the spectra of galaxies of various metallicities, from 5\% to
250\% solar.  
BC99 and BC96 give the spectra of an instantaneous burst 
of star formation (the so called Simple Stellar Population, SSP) that can be
integrated over the time to reproduce an arbitrary SFH.
We used a Salpeter Initial Mass Function (IMF) between 0.1 and 125
M$_{\odot}$. The BC96 model was used mainly because it allows 
to use non solar metallicities.

The FRV97 models use a different approach: starting from some hypothesis
on the efficiency of the star formation processes, they model the entire
life of the galaxy in order to reproduce the observed spectrum. The
resulting SFH for ellipticals is a rapidly declining exponential law giving
most of the star formation in the first Gyr, while for the Sc the SFH is
almost constant over an Hubble time. The
chemical evolution is also computed, but is not used for the final
spectrum which assumes solar metallicity. The Padova isochrones are used, 
together with
the empirical stellar library by Gunn \& Stryker (1983) and
Lan\c con \& Rocca-Volmerange (1992) 
and the IMF by Rana (1992).

The BP97 model has the same approach of BC96, computing the galaxy
integrated spectrum adding up the contribution of SSPs of (possibly)
different metallicities. It
uses the Padova isochrones and the Salpeter IMF between 0.1 and 100
M$_{\odot}$. 
The stellar spectra in the optical are either from Kurucz (1993) or from
Jocoby et al. (1994), while the library by Lan\c con  \& Rocca-Volmerange 
(1996) are used in the near-IR. In this paper we only consider BP97
spectra up to 1.8 $\mu$m because 
above this limit the Kurucz sampling becomes too sparse.

The W94 model provides with single-age stellar population with 
metallicities between 0.01 and 3 times solar. It uses stellar evolution
isochrones by VandenBerg and collaborators and by the Revised Yale
Isochrones (VandenBerg \& Laskarides, 1987; Green et al., 1987),
Salpeter IMF and theoretical stellar spectra by Kurucz (1993) and
Bessell et al. (1994). A particular emphasis was put in studying the
variation with metallicity and age of several optical spectral features.

For demonstration purposes and to better study the intrinsic differences
between the models, for BC99, BC96, W94 and BP97 we used SSP spectra of
solar metallicity. For FRV 
we used both the models of elliptical galaxy they present,
named E13 and E16, corresponding to two different cosmologies.
In all the cases we searched for the
best fitting spectrum by changing the age, the only remaining parameter,
between 5 and 19 Gyr. All the spectra, both observed and modeled, were
normalized to have the same total flux between 0.3 and 2.35 $\mu$m 
and reduced to the same
resolution. The difference between comparing spectra to spectra and 
spectra to broad-band colours is important: both the
absorption features and the continuum shape inside each band are
taken into account and can dominate the results, especially when the
agreement with the colours is good.

To measure the deviation it is not possible to use the standard 
$\chi^2$ because
the fitting points are not totally independent of each other
as the normalization of each band is given by the colour calibration
(see section \ref{sec:calib}). Therefore we need a way to
compare the deviations both to the spectrum noise and to the colour
uncertainties. To do this we compute the square deviation $\Delta^2$
between observed ($o_i$) and
model ($m_i$) spectra in each band:
$$ \Delta^2 = \sum_{i=1}^{N}\frac{(o_i-m_i)^2}{N} $$
where $N$ is the number of spectral points, and divide it for the square
sum $\delta^2$ of the various error contributions. 
The final value $A$ we use as analog to $\chi^2$ is the average of these
ratios in the optical and in three IR bands: 
$$ A = \frac{1}{4}\left[
			\left(\frac{\Delta^2}{\delta^2}\right)_{Op} +
			\left(\frac{\Delta^2}{\delta^2}\right)_J +
       		\left(\frac{\Delta^2}{\delta^2}\right)_H +
       		\left(\frac{\Delta^2}{\delta^2}\right)_K 
		\right]
$$
which has a value of 1 if the deviations are only due to the 
errors in the observed spectra. The quantities $\delta^2$ are difficult
to estimate with precision: from the RMS spread of the spectra (see section
\ref{sec:obs}) and the error on the mean of the colours in table
\ref{tab:colors}, we obtain
$\delta=0.03$ in the optical, J and H and $\delta=0.02$ in K.

In Figure~\ref{fig:fit} we show the resulting values of $A$
as a function of the galaxy age for the four used models and solar
metallicity. Even taking into account the uncertainties on $\delta$,
for all the models an age exists 
which well reproduce the data: BC99 shows the best accord for
ages between 10 and 16 Gyr, W94 select younger ages, between 8 and 9
Gyr, FRV97 reaches the best agreement for older ages, above 16 Gyr. 
The deviation of the BP97 model is always above 1 but steadily
reduces toward older ages, reaching a value of
about 2 at 16 Gyr, the maximum age available for the model.
It should be noted that BC99, BP97 and W94 are single-burst
models with the same IMF and metallicity, 
therefore the difference between them are due to something outside the
user control, as isochrones or stellar spectra. FRV97 is not a single
burst model, about 92\% of the stars are formed in the first Gyr 
(about 97\% in the first 4 Gyr) but the SFR never reaches a null value;
this probably explains at least part of the differences with the other
models and the requirement of very old ages.

In Figure~\ref{fig:bc99} we show the best fitting model, the
BC99 spectrum for an SSP of 12 Gyr
of age, solar metallicity and empirical star spectra. The overall shape 
of the spectrum is very well reproduced: the largest difference are in
the interpolated region between 1 and 1.1$\mu$m where the model trace the
position of the continuum while our interpolation an ``effective'' level taking
into account the absorption lines.
On the contrary, at a closed look it is possible to see that
important discrepancies remains in the details of the absorption lines,
both if theoretical and observed star spectra are used. 
This will be the subject of the next section. 

We also tested the effect of changing the metallicity
by using the BC96, W94 and BP97 models with
metallicities 2.5 times solar, BC96 0.2 solar and W94 0.3 solar.
In all cases no good fit can be reached for any age, as 
the values of $A$ remain above 5.
Models with super-solar metallicities
are too red for $\lambda>8000$\AA, even for young ages (limited to be
higher then 5 Gyr): as an example, all these models predict fluxes in
the H band between 10\% and 20\% higher than observed. Vice versa for
sub-solar metallicities: the H band flux is under predicted of about 20\%.
These discrepancies might be an indication of the real metallicity
of the elliptical galaxies, but it is more probably due
to inadequacy of the non-solar stellar
library or isochrones.  It is beyond the purpose of this paper to solve 
this ambiguity.

\section{The absorption lines}

\begin{table}
\begin{center}
\caption{Feature Identification}
\label{tab:lines}
\begin{tabular}{lcl} 
\hline
$\lambda$&  Main      & Other   \\
($\mu$m) &contribution& species \\
\hline
  1.529  & OH         & CN,TiI         \\
  1.540  & OH         & SiI            \\
  1.558  & $^{12}$CO  & OH             \\
  1.577  & $^{12}$CO  & MgI,FeI        \\
  1.589  & SiI        & OH             \\
  1.598  & $^{12}$CO  & SiI,$^{13}$CO  \\         
  1.606  & OH         &                \\
  1.619  & $^{12}$CO  & OH,CaI         \\
  1.640  & $^{12}$CO  & SiI,[FeII]     \\
  1.652  & $^{13}$CO  & OH             \\
  1.661  & $^{12}$CO  & OH             \\
  1.669  & SiI        & OH             \\
  1.672  & AlI        & HI,$^{12}$CO   \\
  1.677  & AlI        &                \\
  1.689  & OH         & HI,CO          \\
  1.710  & MgI        & CO,OH          \\
  1.723  & OH         & SiI            \\
  1.733  & SiI        & HI             \\
  2.067  & FeI        &                \\
  2.072  & FeI        &                \\
  2.081  & FeI        & SiI            \\
  2.107  & MgI        & H$_2$O,SiI     \\
  2.117  & AlI        & H$_2$,MgI,FeI  \\
  2.136  & SiI        &                \\
  2.146  & MgI        & NaI,SiI,CaII   \\
  2.166  & Br$\gamma$ & VI             \\
  2.173  & ScI        & FeI            \\
  2.179  & TiI        & SiI,FeI        \\
  2.189  & SiI        & TiI,FeI        \\
  2.208  & NaI        & ScI,TiI,VI,FeI,SiI \\
  2.226  & FeI        & ScI,TiI        \\
  2.239  & FeI        & ScI            \\
  2.248  & FeI        & VI,TiI         \\
  2.263  & CaI        & ScI,TiI,FeI,SI \\
  2.281  & MgI        & CaI,FeI,SI,HF  \\
  2.294  & $^{12}$CO  & TiI            \\
  2.323  & $^{12}$CO  &                \\
  2.345  & $^{13}$CO  &                \\
\hline
\end{tabular}
\end{center}
\end{table}

The use of our spectra to study the absorption line is hampered by 
the low resolution, both because
faint lines are below the detection threshold and because many
lines are blended together. Nevertheless, many features can be easily 
seen and measured with great accuracy, even if
in some cases their identification may not be unique.
For the following analysis we have only used the spectra in H and K
because a series of facts make our J spectra less useful
to study the features: they have the lowest resolution, their SNR is 
lower than that in the other bands, the features are intrinsically less
prominent, and the star classification in this band is still in its
infancy (see, for example, Wallace et al., 2000).

To further increase the SNR we decided to use the average of 
the spectra of E, S0 and Sa because they show
no significant difference in the absorption lines, meaning that
temperature, luminosity and metallicity of the dominant 
stellar populations are similar. 
We will indicate with ES the resulting average spectrum.

We identified the lines (see Table~\ref{tab:lines})
by comparing the features in our spectra 
to those identified in the late type stars.
The literature of reference is now rich enough
(e.g., Origlia et al., 1993; Oliva et al., 1995, 1998; 
Meyer et al, 1998, hereafter M98; Kleinmann \& Hall, 1986, hereafter KH86; 
Vanzi \& Rieke 1997; Wallace et al, 1996, 1997) 
to make the identifications reliable, but the dominant contribution to
some features due to blending of many lines is sometime uncertain. 
In these cases, the synthetic spectra constructed
specifically for line identification in the H band 
by Origlia et al. (1993) were of particular help,
while in the K band the high-resolution spectra by Wallace et al. (1996)
were taken into particular account.

\begin{figure}
\centerline{\psfig{figure=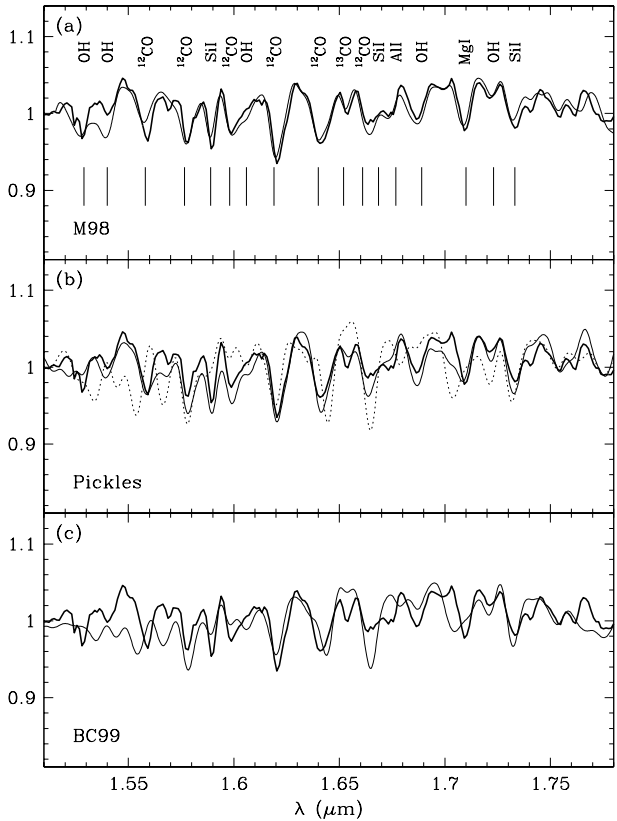,width=9cm}}
\caption{
Line identifications and comparison of the observed H-band 
spectrum of the early-type galaxies 
(thick line) normalized to its continuum 
to several spectral libraries (thin lines) reduced to the
same resolution. 
Panel (a): the galaxy spectrum is compared to a K5-giant star 
from the libraries by M98, showing a very good agreement. 
Panel (b): the spectra of a K5 (solid) and a MO giant (dotted) 
from Pickles (1998) are overplotted to the galaxy data. Even
if the two stars are similar, their spectra in this library are very
different.
Panel (c): the SSP model from BC99 at 10 Gyr with empirical stellar libraries
(see previous section and Figure~\ref{fig:bc99}) is compared to the
data.
}
\label{fig:Hstars}
\end{figure}

\begin{figure}
\centerline{\psfig{figure=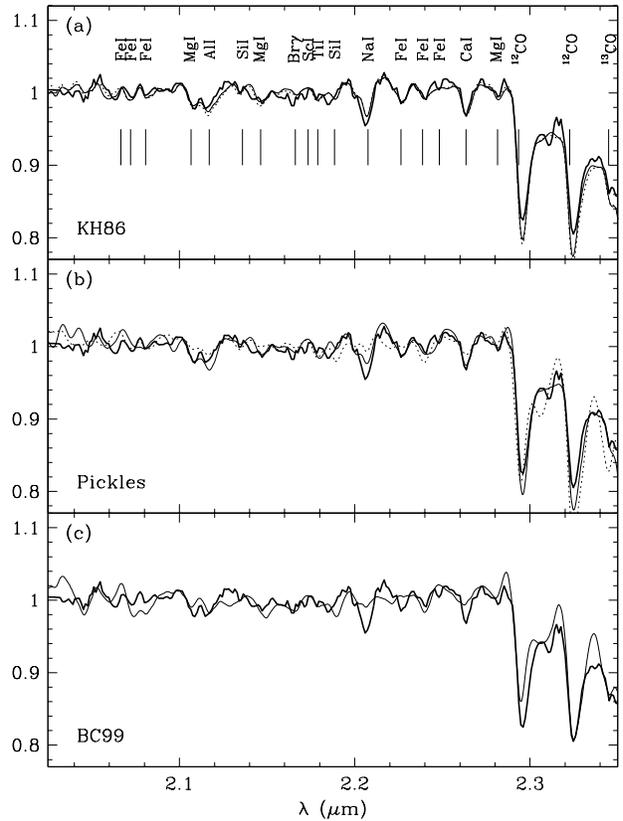,width=9cm}}
\caption{
As the previous figure, but in the K band.
The K5III star spectrum in panel (a) is from KH86.
}
\label{fig:Kstars}
\end{figure}

The first useful results can be derived by qualitatively comparing
the H- and K-band parts of the ES spectrum to various stellar spectra
libraries and to the results of the spectrophotometric models. 
This comparison is shown in Figures~\ref{fig:Hstars} and \ref{fig:Kstars}
where all the spectra were normalized to a fitted continuum
and reduced to the same resolution of the observed data.

In panel (a) of Figure~\ref{fig:Hstars}
a K5III star from M98 is overplotted to the data. The agreement is 
very close and basically all the
features are well reproduced: as expected, cold giant stars dominate 
the emission at 1.6 $\mu$m and the observed spectra can be reproduced even
with a single star spectrum 
Analogous excellent agreement is obtained using spectra
from the Wallance \& Hinkle (1996) library (not shown) which are limited to a
narrower wavelength range.

The following panel 
shows the spectrum of two stars, a K5III and a M0III, from the 
more recent stellar libraries by Pickles (1998), library used in several
spectrophotometric models. 
The former star, the same as panel (a), shows a spectrum similar to our
data, even if the agreement is not as accurate as for the KH86 spectrum.
On the contrary, the M0III star shows a very different behavior, as also other
stars of similar spectroscopic and luminosity class. These
differences are not explained by the small temperature difference between the
K5 and M0 stars (about 160 $^o$K, M98) and such a behavior is not seen in other
libraries, as M98 or Wallance \& Hinkle (1996). 
The differences are probably due to the use of several  
different sources in the derivation of the Pickles (1998) library. In
particular Pickles bases the near-IR part of its spectral on the 
the Lan\c con \& Rocca-Volmerange (1996) library because it covers 
a wide wavelength range, between 1.43 and 2.5 $\mu$m, with 
accurate flux calibration. Unfortunately, some features of the spectra 
of this library seem very variable from one spectral type to the next ones and
very different from other libraries (as M98 and KH86) and from our spectra.
Therefore they probably are observational artifacts.
As a consequence, any spectrophotometric model based on this library will 
not be able to accurately reproduce the features in near-IR galaxy spectra.
This is shown in the following (c) panel: here the observed spectrum is
compared to the BC99 SSP galaxy model
at the age of the best fit in Figure~\ref{fig:fit}. 
This model is the only one in our sample having a resolution in
the near-IR high enough to allow a meaningful comparison of the 
spectral features.
Even if the general shape of the continuum is well reproduced, as shown in 
Figure~\ref{fig:fit}, the agreement with the spectra features is generally
poor: as an example, the deep absorption lines predicted by the models at 
1.665 and 1.705 $\mu$m are not present in the observed data. 

Same conclusions follow from the spectra in the K band shown in
Figure~\ref{fig:Kstars}: the K5III star from KH86 in panel (a) shows a very
good agreement with the ES spectrum. Only the CO bands at 2.29 and 2.32 $\mu$m
are deeper in the star spectrum than in the ES one, because of either
metallicity effects, or the contribution of stars warmer than K5 or on the
main sequence, or differences in the fitted continuum level above 2.28 $\mu$m. 
Analogous excellent agreement is obtained with the K5III star spectrum 
from the Wallance \& Hinkle (1996) library (not shown). 

In panel (b) and (c) the Pickles (1998) and BC99 spectra are shown: again, some
features are not reproduced. For example, 
the prominent NaI feature at
2.21 $\mu$m  is almost completely absent from the M0III star and from the
model spectra. Also the CaI feature at 2.26 $\mu$m appears deeper in the
observed data than in the model spectrum.
Given that the general shape of the continuum is well reproduced by the model, 
(see the previous section) the problem must be in the used stellar library.  

To summarize, libraries of stellar spectra exist which reproduce the 
observed galaxy
spectra, while others show very different features. Spectrophotometric models
based on the second set of spectra might not be able to reproduce the observed
features. Our spectra will make it possible to tune the 
models, especially when new and more extended libraries of
near-IR stellar spectra will be available
(see, for example, Ivanov et al, 1999; Mouhcine \& Lan\c con, 1999).

\section{The dominant stellar population}

As discussed in the previous section, 
indications on the dominant stellar population 
can be easily derived even by simply comparing the observed data with the
stellar spectra. 
More quantitative results can be obtained by
comparing the absorption lines. 

Several authors have used this method to classify the
stars by their near-IR spectra (see references below).  
Usually two sets of lines are considered
to derive both the spectral and the luminosity class,
the first set depending both on temperature and gravity, 
while the second one on temperature only.
The recipes used for star classification can also be
applied to the observed galaxy spectra to derive indications on the 
the dominant stellar population. 

We took into account two schemes of classification.
They are both based on the CO(2,0) band head at 2.29 $\mu$m whose
intensity depends both on temperature and luminosity and can be used to
study the stellar populations in galaxies or in dusty regions
(e.g., Armus et al., 1995; James \& Seigar, 1999; Mobasher \& James,
2000).
The two methods use different thermometers: the first one, 
introduced by  KH86 and refined by Ramirez et al. (1997), is based on the
Equivalent Width (EW) of NaI 2.21$\mu$m and CaI 2.26$\mu$m
which depend strongly on temperature but not on luminosity.
In the second one, based on Origlia et al. (1993), the temperature is measured
by the ratio of the two features at 1.62 and 1.59 $\mu$m. 
In both cases, the results
for the galaxies are compared with those for the stars of known
properties.

To apply these methods, we measured the EW of the lines 
with respect to a local continuum obtained
by fitting a straight line to the "clear" parts of the spectrum.
The spectral region used for both the line and the continuum 
are listed in Table~\ref{tab:features}.  The same procedure was used to
measure the EWs of the stars of the M98 and KH86. Because of the 
lower resolution of our spectra, our definitions are a little different
from those by, for example, KH86 or Origlia et al. (1993), and the
measured values of the EWs are consequently different.
The errors on these quantities were estimated by taking into account 
both the noise of the
spectra and the uncertainties in the continuum level. The latter
contribution is usually the dominant one, especially in H where many
features are present and it is difficult to define the continuum level. 
The results for the lines used in the classification
are shown in Table~\ref{tab:ew1} for the average ES spectra and
in Table~\ref{tab:ew2} for the individual galaxies not used for the
templates. 
It should be noted that the error on the 1.62/1.59 ratio shown in the
table is smaller than the
error on each single line because the same continuum fit is used:
the uncertainties on the continuum level strongly affect
the measure of the EW of each line, but partially cancel out from their
ratio.

\begin{table}
\begin{center}
\caption{Feature definition}
\label{tab:features}
\begin{tabular}{ccc} 
\hline
Feature        &     Line    & Continuum \\ 
               &  ($\mu m$ ) & ($\mu m$) \\ 
\hline
 1.59          & 1.586-1.594 & 1.572-1.574, 1.608-1.613 \\  
               &             & 1.628-1.633              \\  
 1.62          & 1.616-1.629 & 1.572-1.574, 1.608-1.613 \\  
               &             & 1.628-1.633              \\  
NaI            & 2.204-2.211 & 2.191-2.197, 2.213-2.217 \\ 
CaI            & 2.258-2.269 & 2.245-2.256, 2.270-2.272 \\ 
$^{12}$CO(2,0) & 2.289-2.302 & 2.270-2.272, 2.275-2.278 \\ 
	           &             & 2.282-2.286, 2.288-2.289 \\ 
\hline
\end{tabular}
\end{center}
\end{table}

\begin{table}
\begin{center}
\caption{Value of the EW (\AA) measured on the templates, and their errors}
\label{tab:ew1}
\begin{tabular}{cccc} 
\hline
Feature        & E+S0+Sa   &   Sb      & Sc      \\
\\
\hline
1.59           & 3.9 (0.8) & 3.7 (0.9) & 3.3 (0.9) \\
1.62           & 6.0 (0.7) & 5.6 (0.8) & 5.7 (0.8) \\
1.59/1.62      & 1.5 (0.1) & 1.5 (0.2) & 1.7 (0.2) \\
$^{12}$CO(2,0) &14.6 (1.7) &15.8 (1.7) &15.6 (1.7) \\
CaI            & 2.6 (1.1) & 2.7 (1.1) & 2.2 (1.1) \\
NaI            & 3.4 (0.7) & 4.2 (0.7) & 3.4 (0.7) \\
\hline
\end{tabular}
\end{center}
\end{table}

\begin{table}
\begin{center}
\caption{Value of the EW (\AA) measured for the active galaxies, 
    and their errors}
\label{tab:ew2}
\begin{tabular}{cccc} 
\hline
               &  NGC2798   & NGC1569     & NGC4449    \\ 
\\
\hline
$^{12}$CO(2,0) & 11.8 (2.4) & 16.7 (2.3)  & 13.8 (2.9) \\ 
NaI+CaI        & $<3.7$     & $<4.2$      & $<5.8$     \\ 
\hline
\end{tabular}
\end{center}
\end{table}

\begin{figure}
\centerline{\psfig{figure=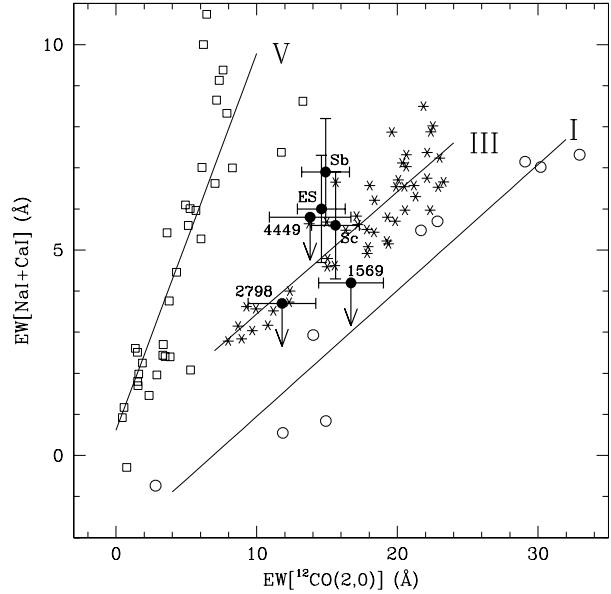,width=9cm}}
\caption{The sum of the EWs of NaI (2.21 $\mu$m) and CaI
(2.26 $\mu$m) is plotted versus the EW of the $^{12}$CO(2,0) band at 2.29
$\mu$m. 
The solid dots with the error bars are our galaxies, while
dwarf, giant and supergiant stars are plotted as
empty squares, stars and empty circle respectively (see text for
references). The lines are
linear fit to the stellar data, with the indication of the luminosity class.
}
\label{fig:righeK}
\end{figure}

\begin{figure}
\centerline{\psfig{figure=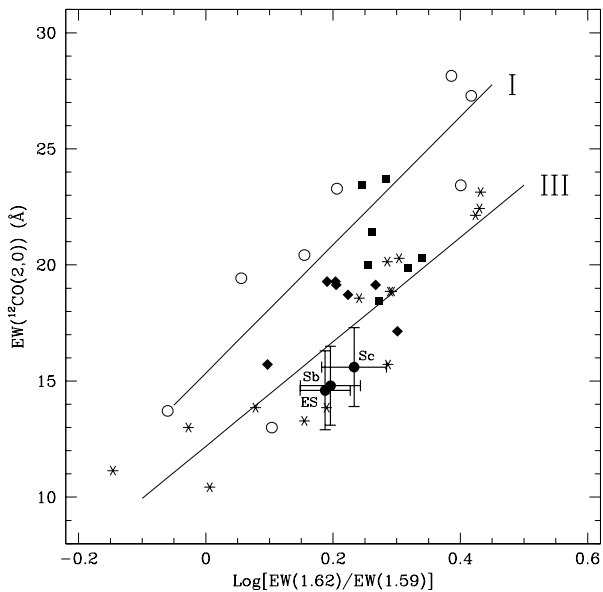,width=9cm}}
\caption{EW of the $^{12}$CO(2,0) feature at 2.29$\mu$m vs. the 
ratio EW(1.62)/EW(1.59). Giants and supergiants stars (see text for
references) are plotted by
empty circles and stars respectively. 
The galaxies observed by Origlia et al. (1993) are plotted by solid diamonds 
if quiescent ellipticals and spirals, and solid squares if HII galaxies.
Solid circles with error bars 1are the galaxies of this study. The lines
are the linear fit to the stellar data.
}
\label{fig:righeH}
\end{figure}

In Figure~\ref{fig:righeK} we present the classification based on
the features in the K band studied by KH86 and Ramirez et al. (1997).
This figure differs from both works because the feature definition is
different (because of the lower resolution) and because we don't take
into account the EW of Br$\gamma$ which cannot be reliably measured.
As expected, the results show that in all the cases the main role 
is played by class III stars, possibly with some
contribution from dwarf stars. With the present data it does not seem
possible to measure this contribution and therefore constraint the ratio
between dwarf and giant stars, but this could be done with higher
resolution observations.
The spectra of the active galaxies (see Figure~\ref{fig:active}) are too
noisy to measure the EWs of NaI and CaI, and only upper limits can be
derived. As shown in Figure~\ref{fig:righeK}, these upper limits 
put two of these galaxies in the region between class I and class III
stars, suggesting that the contribution supergiants
might be important in these galaxies.

In Figure~\ref{fig:righeH} we apply the second method based on the CO
feature at 2.29$\mu$m and on the ratio of the 1.59 and 1.62$\mu$m in the
H band. This method is based on the work by Origlia et al (1993) who 
showed that this ratio can be use as a thermometer in case of giants and
supergiants stars. Also in this case the line ratio is consistent with
the class III stars.\\

A discussion about the metallicity of ellipticals and early spirals is
beyond the scope of this paper. Detail discussions about the information
of metallicity and element ratios based on optical data can be found, for
example, in Worthey (1998) and Trager et al. (2000).  
Origlia et al. (1997) have shown that moderate-resolution near-IR spectra can
also be used to measure the metallicity of the systems.
Their recipe is based on the 1.62$\mu$m feature which 
is shown to depend on the metallicity index [Fe/H]
and, more weakly, on the carbon depletion factor [C/Fe]. 
It also depends on the microturbolent velocity $\xi$ of the dominant 
stellar populations because this quantity
influences the strength of the
saturated molecular lines. By observing globular clusters of known
metallicity, Origlia et al. (1997) show that this metallicity scale gives
errors smaller that 0.3 dex. These authors have also observed a limited
number of elliptical galaxies. The few metallicity estimates published
for these objects are based on the Mg2 index (see, for example,
Davies et al., 1987) and on the Mg2-metallicity calibration by
Casuso et al. (1996). While the Mg2-derived value of [Fe/H] of the observed
galaxies was solar or slightly supersolar, between +0.08 and +0.21, 
the values based on the
near-IR lines turned out to be subsolar, between -0.28 and -0.46. This
can be explained either by the presence of a corresponding Magnesium
enhancement of about [Mg/Fe]$\simeq$0.5, as proposed by several authors
(see Origlia et al., 1997 for details), or by an anticorrelation between
[C/Fe] and [Fe/H], the more metallic objects being more carbon depleted, as
observed in the globular clusters. More recently, several authors (Trager et
al., 2000; Worthey, 1998) would rather explain the non-solar [Mg/Fe] with
a depletion of the iron-peak elements. 

Despite the large uncertainties on the right value of [C/Fe]
and on the resulting metallicity, we have applied this method to our
early-type galaxies (ES spectrum, see table~\ref{tab:ew1}).
By iteratively solving the eq. 2 by Origlia et al. (1997) 
we obtain a mean microturbolent velocity $\xi$ of about 3.1 $km/sec$,
in agreement with the values between 2.3 and 3.0 in Origlia et al. (1997).
By using the 'standard' value of [C/Fe]=$-$0.3 suggested by these authors
and derived for the Magellanic Clouds clusters by Oliva \& Origlia (1998), 
the metallicity derived from their eq. 1b is [Fe/H]$\simeq-0.20$, reducing to
[Fe/H]$\simeq-0.4$ for [C/Fe]=0.0.
The error on this quantity is difficult to estimate and is probably
dominated by the uncertainties on [C/Fe], but this result can be used as an
indication of sub-solar abundance of iron in the early-type galaxies.

\section{Summary}

The near-IR spectra of 28 nearby galaxies were used to define the
``average'' J, H and K spectra of normal galaxies along the 
Hubble diagram between E and Sc. The target objects and the aperture 
were chosen to be similar to those by K96 in order to merge the two data
sets and obtain self-consistent spectra between 0.1 and
2.4 $\mu$m. The average effective colours of the galaxies of the various
classes were used to calibrate the final spectra. The overall precision
is about 2\%.

The spectra were used to compute the k-corrections in the near-IR bands
and to check the predictions of several spectrophotometric
models of galaxy evolution. We find that most model can accurately reproduce
the observed continuum shape of the elliptical galaxies, if solar
metallicities are assumed. In this case the derived SSP ages at above 8 Gyr.
Nevertheless, large differences are found 
between the models, for example in terms of the derived age of the
dominant stellar population.

On the other hand, larger discrepancies are found between the observed 
spectral features 
and the predictions of the models. We note that the spectra of cool giant
stars from some accurate libraries (as KH86 and M98)
can reproduce the observed features
in great detail, and conclude that the disagreement
is probably due to the stellar libraries used in the models.

The spectral features are also used to study the stellar populations 
by using several stellar classification recipes based on near-IR
absorption lines, some of them applied to galaxies for the first time.
The dominant contribution to the near-IR emission of all the galaxies
between E and Sc is confirmed to be due to giant stars of spectral type
about K5, with solar or slightly subsolar metallicity.\\

The final spectra for each galaxy type, the relative k-corrections
(both in the Bessel \& Brett and in the CIT systems)
and the profiles of the used filters
can be downloaded from the web site 
\underline{www.arcetri.astro.it/$\sim$filippo/spectra}. 
In some cases it could be useful to simulate the spectra of galaxies 
with colours different from those in Table~\ref{tab:colors}, for example
to reproduce individual galaxies or match total colours. 
In the same web site it is possible to
define a new set of colours, compute the corresponding spectra
and download the results.

\section{Acknowledgments}

We are grateful to Tino Oliva, Livia Origlia and Valentin Ivanov,
for enlightening discussions about near-IR spectra and stellar
classification, and to Stephan Charlot and Gustavo Bruzual for having
provided us a recent version of their model. 
We are also grateful to the TIRGO staff for the support during
the observations. LP acknowledges a partial support from the ASI grant 
ASI-ARS99-44.


\end{document}